\documentclass[runningheads]{llncs} 

\usepackage[utf8]{inputenc}
\usepackage{hyperref}
\usepackage{caption}

\usepackage{todonotes}
\usepackage[binary-units=true]{siunitx}
\usepackage{float}

\usepackage{booktabs}
\usepackage{multirow}
\usepackage{subcaption}
\usepackage{tabularx}

\DeclareSIUnit\flop{FLOP}
\DeclareSIUnit[per-mode=symbol]\floppersec{\flop\per\second}

\title{JUWELS Booster -- A Supercomputer for Large-Scale AI Research}
\author{Stefan~Kesselheim\inst{1}\thanks{equal contribution},
        Andreas~Herten\inst{1}\printfnsymbol{1},
        Kai~Krajsek\inst{1}\printfnsymbol{1},
        Jan~Ebert\inst{1}\printfnsymbol{1},
        Jenia~Jitsev\inst{1}\printfnsymbol{1},
        Mehdi~Cherti\inst{1}\printfnsymbol{1},
        Michael~Langguth\inst{1}\printfnsymbol{1},
        Bing~Gong\inst{1}\printfnsymbol{1}, 
        Scarlet~Stadtler\inst{1}\printfnsymbol{1},
        Amirpasha~Mozaffari\inst{1}\printfnsymbol{1},
        Gabriele~Cavallaro\inst{1}\printfnsymbol{1},
        Rocco~Sedona\inst{1,2}\printfnsymbol{1},
        Alexander~Schug\inst{1,3}\printfnsymbol{1},
        Alexandre~Strube\inst{1},
        Roshni~Kamath\inst{1},
        Martin~G.~Schultz\inst{1},
        Morris~Riedel\inst{1,2},
        Thomas~Lippert\inst{1}
        }
        
\authorrunning{S. Kesselheim et al.}

\institute{Jülich Supercomputing Centre, Forschungszentrum Jülich GmbH, Germany, contact \email{<n>.<surname>@fz-juelich.de}
\and
School of Engineering and Natural Sciences,\\ University of Iceland, Reykjavik, Iceland
\and
University of Duisburg-Essen, Germany
}

\makeatletter
\newcommand{\printfnsymbol}[1]{%
  \textsuperscript{\@fnsymbol{#1}}%
}
\makeatother

\begin{document}

\maketitle

\begin{abstract}
In this article, we present JUWELS Booster, a recently commissioned high-performance computing system at the Jülich Supercomputing Center. With its system architecture, most importantly its large number of powerful Graphics Processing Units (GPUs) and its fast interconnect via InfiniBand, it is an ideal machine for large-scale Artificial Intelligence (AI) research and applications. We detail its system architecture, parallel, distributed model training, and benchmarks indicating its outstanding performance. We 
exemplify its potential for research application by presenting large-scale AI research highlights from various scientific fields that require such a facility.

\end{abstract}

\section{Introduction}
\label{section:introduction}
In recent years, deep learning methods brought up radical transformation across various disciplines of technology and science, sparking interest in the previously academic field of Artificial Intelligence (AI), and extending it far beyond academia. Progress is notified and valued in mainstream and business media and widely discussed.  Without disparaging many other ingenious contributions, it can be said that a single paper started the boom: in 2012, Alex Krizhevsky, Ilya Sutskever and Geoffrey Hinton had shown that a deep neural network outperforms the state-of-the-art in image classification in the ImageNet competition, by a large margin ~\cite{krizhevsky2012imagenet}. 
Since then, interest has skyrocketed, and the number of publications on AI has grown exponentially~\cite{zhang2021ai}.

A vital ingredient of this success was the availability of computational resources for AlexNet: two NVIDIA GeForce GTX 580 graphics processing units (GPUs).
Since then, not only the cumulated amount of compute has increased tremendously, but also the computational resources needed by AI models. Amodei et.\ al observed a \num{300000}-fold increase in six years of the computational effort required for the largest models~\cite{amodei2018ai}. This trend extends over all fields of machine learning (ML), ranging from Computer Vision (CV), over Natural Language Processing (NLP) to Reinforcement learning (RL).

A prominent example with a lot of media echo is the GPT-3 model, a transformer-based NLP architecture consisting of 175 billion parameters trained on a corpus of \SI{45}{\tera\byte} of text~\cite{brown2020language}. The impressive NLP model is capable of performing ``in-context learning'', i.e.\ it can pick up novel tasks at inference time without re-training by just providing an input that verbally describes the task or gives the context of the desired outcome.

In the CV field, the BigTransfer~(BiT) work follows similar lines. A very large model is pre-trained on an extremely large dataset: up to 300 million images, allowing for fine-tuning the model in an extremely data-efficient way for downstream tasks~\cite{Kolesnikov2020}. This idea is further detailed in Section ~\ref{sec:transfer}. Self-supervised approaches make it possible to also use data that has not been manually annotated for pre-training~\cite{chen2020big}. Such approaches heavily rely on large models and extensive training procedures but have the potential to reach learning from only a few examples. An even greater source of visual data enclosing also the temporal dimension are videos as shown e.g.\ in Ref.~\cite{orhan2020self}. 


While the previous examples involve the private sector, also in the academic world, large scale training have been performed. Kurth et.\ al trained an ML-based large scale climate model on the Piz Daint and Summit supercomputers, reaching 1~EFlop/s peak in FP16~\cite{kurth2018exascale}. Other typical situations in science are inverse problems, in which an accurate forward simulation model is available, but the practical task is the inference of parameters. In an example of Laanait et al., electron densities were inferred from diffraction patterns by an ML model, a computation that involved a training data set of \SI{500}{\tera\byte} and up \num{27600}~GPUs on Summit~\cite{laanait2019exascale}. A further example is the evolutionary search of network architectures in cancer research~\cite{patton2019exascale}. Here, the objective was to find an architecture that could analyze microscopic tissue slides to support cancer diagnosis and that is suitable for usage in a desktop PC application, requiring an extensive search for the optimal model. 

Reproducibility is an indispensable factor for AI research~\cite{stodden2016enhancing}. While many academic papers are accompanied with dedicated repositories, reproducing the largest models from the industry sector are especially difficult, for neither of the examples GPT-3, AlphaZero, AlphaFold or DALL-E the full source code or trained models have been released. Initiatives dedicated to reproducing such results can make very important contributions to the field, yet require considerable computational resources \footnote{See e.g.\  \url{https://github.com/EleutherAI/the-pile}}. 

To support these trends, a new supercomputer has been installed at the Jülich Supercomputing Centre~(JSC) for Forschungszentrum Jülich. The machine is installed as a booster module for the modular supercomputer JUWELS (Jülich Wizard for European Leadership Science) and referred to as JUWELS Booster. Almost 4000 high-performance GPUs render it both one of the fastest and most energy efficient computers in the world, currently ranked no.~7 in the Top500 list and no.~3 in the Green500 list, as well as the fastest supercomputer in Europe. It is intended as a machine that not only supports AI research, but can act as a substrate for a blooming AI community. 

In this paper, we will explain the rationale behind the system design, indicate its technological capabilities, and demonstrate its potential by showing several research highlights from Forschungszentrum Jülich. This paper is structured as follows. After the introduction in Section~\ref{section:introduction}, we present JUWELS' ecosystem in Section~\ref{section:ecosystem}. Here, we discuss its architecture, parallel model training techniques and ML benchmark results. In Section~\ref{section:largescale}, we present four examples of large scale AI applications. This includes large scale transfer learning, deep-learning driven weather forecast, multispectral remote sensing image classification, and RNA structure analysis. We conclude with a summary and an outlook.

\section{JUWELS Booster System}
\label{section:ecosystem}

\subsection{JSC Supercomputer Ecosystem}
The Jülich Supercomputing Centre operates different supercomputers of various sizes. The two largest ones are JURECA and JUWELS. JUWELS is a \emph{Gauss Centre for Supercomputing} Tier~0/1 machine~\cite{juwels2019}, currently consisting of two modules: JUWELS Cluster and JUWELS Booster. While the JUWELS Cluster module provides general-purpose computational resources with more than 2300 compute nodes based on Intel Skylake CPUs, JUWELS Booster is the system's highly scalable module, leveraging GPUs to provide computing performance. Both modules are combined through their network fabric and file system and can be used together, by heterogeneous jobs, through a tight integration via the workload manager.

\subsection{JUWELS Booster}

Commissioned in 2020, JUWELS Booster features the latest GPU, CPU, and network technology available. \num{936} compute nodes host four GPUs each, providing access to \num{3744}~GPUs. The installed GPUs are NVIDIA A100 Tensor Core GPUs~(\SI{40}{\giga\byte}), providing \SI{19.5}{\tera\floppersec} of $\text{FP64}_\text{TC}$ computing performance each. The GPUs are hosted by AMD EPYC 7402 CPUs with $2\times 24$~cores (SMT-2) per node. Each node is equipped with \SI{512}{\giga\byte} of RAM.

The network of JUWELS Booster is based on Mellanox HDR200 InfiniBand, with four Mellanox ConnectX~6 devices per node, each providing \SI{200}{\giga\bit\per\second} of bandwidth per direction. The network is designed as a DragonFly\texttt{+} network; the nodes are aligned in sets of~48 in a local switch group (\emph{cell}). While nodes in a cell are tightly connected as a full fat-tree with two levels of switches, each cell connects to the other cell with 10~links. The resulting total bi-section bandwidth is \SI{400}{\tera\bit\per\second} between the cells. Dedicated network links are available to connect to JUWELS Cluster. These links also provide access to a highly-parallel, flash-based file system with \SI{1400}{\giga\byte\per\second} peak bandwidth. The storage cluster, JUST, can be reached with a peak of \SI{400}{\giga\byte\per\second} bandwidth via gateway nodes.


The NVIDIA A100 GPUs installed into JUWELS Booster provide different computing performance depending on the precision used. Within the \SI{400}{\watt}~TDP, the following peak performance is available: \SI{9.7}{\tera\floppersec} ($\text{FP64}$), \SI{19.5}{\tera\floppersec} $\text{FP64}_\text{TC}$ and $\text{FP32}$, \SI{78}{\tera\floppersec} $\text{FP16}$, \SI{156}{\tera\floppersec} $\text{TF32}_\text{TC}$, \SI{312}{\tera\floppersec} $\text{FP16}_\text{TC}$, where \emph{TC} denotes the usage of Tensor Cores. With respect to the $\text{FP64}$ Tensor Cores, an excellent peak efficiency of \SI{48,75}{\giga\floppersec\per\watt} can be reached. Indeed, JUWELS Booster ranks highest in the Green500 list of November 2020 as the most energy-efficient supercomputer within the first 100 places of the Top500 with \SI{25}{\giga\floppersec\per\watt}.

The software stack on JUWELS is managed via EasyBuild and accessed via environment modules, enabling finely-tuned, homogeneous software environments across the modules of the system. In addition, containers are supported via Singularity, either by using pre-made containers from registries, or by building containers directly from the system through a dedicated container build service.

\subsection{Distributed Model Training on JUWELS Booster}
The execution of efficient AI algorithms on JUWELS Booster requires different levels of parallelism: At the level of basic numerical operations, at the level of parallelization of deep learning models, and at the level in the training procedure of the deep learning models. 

Modern deep learning models transform $n$-dimensional tensors by applying element-wise operations, e.g.\ activation functions, convolution operations, or matrix multiplication in fully connected layers. While element-wise operations are embarrassingly parallelizable, convolution operations and matrix multiplication require special, sophisticated parallelization strategies. The development of parallel matrix operations and convolutions that can themselves be reduced to matrix operations is mature and results in highly optimized libraries such as MKL~\cite{MKL09}, cuBLAS~\cite{CUBLAS17}, and cuDNN~\cite{chetlur2014cudnn}. JUWELS Booster's GPUs contain Tensor Cores, hardware that specializes in matrix multiplications in the context of AI. In conjunction with cuBLAS, it is possible to apply efficient convolution operations and matrix multiplications in GPUs. Depending on the dimension of the input tensor and its data type, the most efficient algorithm is automatically selected by the cuBLAS library. While it is not necessary to optimize these operations manually, AI applications can be tuned by choosing optimal input dimensions and data types, i.e.\ Tensor Cores work most efficiently when the data dimension is divisible by a certain number depending on the data type. Reducing the precision of the data type, for example from FP32 to FP16, can also lead to a significant speed-up.  The deep learning frameworks installed on JUWELS Booster (TensorFlow ~\cite{tensorflow2015-whitepaper} and PyTorch~\cite{paszke2019pytorch}) provide modules (e.g.\ \texttt{torch.cuda.amp} in PyTorch) to automatically reduce precision for operations where precision is a minor concern. While operator-level parallelization is performed on a single GPU, model-level parallelization enables usage of all \num{3744}~GPUs of the JUWELS Booster system. A common method for parallelizing the training of neural networks is data parallelism. This involves replicating the model on several GPUs, each of which is trained with a different batch of the training data. Apart from batch normalization~\cite{pmlr-v37-ioffe15}, all operations in modern deep learning models can be performed independently on the different computing devices. After calculating the gradient of the model parameters, the gradients are averaged over the different replicants, which effectively gives the same result as training a model on a large batch~-- the combination of all distributed data batches. The averaging of the gradients as well as the batch normalization operation require collective communication across different GPUs, and can become a bottleneck when scaling the training process. To overcome these problems, one usually changes the training procedure in a distributed environment. For example for batch normalization, one computes the batch statistics only on a subset of the parallel training batches involved, which has even been shown to be advantageous over computing over exceptionally large total batches~\cite{DBLP:journals/corr/GoyalDGNWKTJH17}. Collective communication can be accelerated by compressing the gradients before averaging~\cite{dettmers2015approximations,NEURIPS2019_d9fbed9d,Agarwal2021OnTU}. The JUWELS Booster software stack provides Horovod~\cite{Sergeev2018} as a data-parallel framework with NCCL\footnote{https://docs.nvidia.com/deeplearning/nccl/index.html} as a communication framework that works with the supported deep learning frameworks TensorFlow and PyTorch, and comes with built-in FP16 gradient compression. Horovod is applied for the parallel training results shown in this paper. 
JSC also supports HeAT~\cite{heat20}, a general distributed tensor framework for high-performance data analytics. The communication framework of HeAT provides the full MPI~\cite{mpi2015mpi} functionality and  shows a significantly speedup to the comparable library Dask~\cite{rocklin2015dask}.
Large deep learning models may not fit on a single computational device, requiring an extension of the purely data-parallel approach to model parallelism~\cite{Muller94} or pipelining~\cite{Deng12}, which in turn requires automatic differentiation~(AD) across different computational devices. AD is a critical feature of modern deep learning libraries, but most libraries\footnote{PyTorch allows AD for distributing tensors across computational devices based on the remote procedure call~(RPC) protocol~\cite{birrell84rpc}. However, the RPC framework does not compete with communication frameworks like NCCL or MPI with respect to performance.} support AD only on a single node, and no deep learning library supports AD with MPI requiring extensions. JSC supports  DeepSpeed~\cite{ren2021}, a cutting-edge deep learning optimisation library that supports data and model parallelisation as well as pipelining schemes or all in combination  by a lightweight wrapper around PyTorch. In addition JSC supports MPI4Torch\footnote{https://github.com/helmholtz-analytics/mpi4torch} allowing to write PyTorch code directly in distributed environments.
The last level of parallelization considers the parallelization of the training procedure, which includes ensemble learning~\cite{DBLP:journals/corr/LeePCCB15}, model averaging~\cite{NIPS2015_d18f655c}, but also hyperparameter search~\cite{Lorenzo2017HyperparameterSI}, as well as model architecture search~\cite{2017arXiv171100436L}. JSC supports training parallelization with the framework L2L~\cite{subramoney18}. 


\subsection{Benchmark Results}

\begin{figure}[!tb]
    \centering
    \includegraphics[width=\columnwidth]{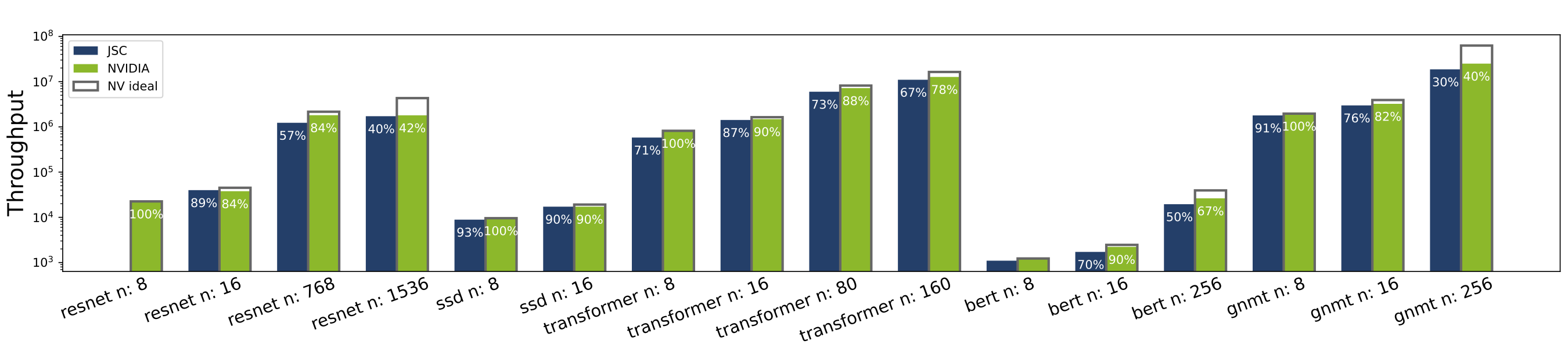}
    \caption{Benchmark results of a subset of MLPerf training~v0.7 runs for different tasks and different number of GPUs~($n$). Our results shown in blue, NVIDIA's in green. Empty bars indicate the throughput for ideal scaling conditions on NVIDIA's results. For each run the efficiency normalized by NVIDIA's single-node result is indicated in percent.}
    \label{fig:benchmark}
\end{figure}

We investigate the capability of JUWELS Booster to achieve nominal performance under practical machine learning conditions by running the MLPerf training benchmark following the~v0.7 submission conditions~\cite{mattson2020mlperf}. As our machine shares key features with NVIDIA's Selene machine, we also run their submission code. To enable comparison, the number of nodes is doubled, as Selene features eight GPUs per node over JUWELS Booster's four. Containers are updated as well. The MLPerf training benchmark measures the time-to-accuracy. This was carefully optimized by NVIDIA by hyperparameter optimization. As we only re-run their benchmarks and do not optimize the hyperparameters, we instead report the throughput in images/second for the tasks \textit{resnet} and \textit{ssd}, words per second for the tasks \textit{transformer} and \textit{gnmt}, and in sequences per second for ``bert''. As indicated in Fig.~\ref{fig:benchmark}, we are  able to closely reproduce NVIDIA's results. Details of our implementation are published on our GitLab server\footnote{\url{https://gitlab.version.fz-juelich.de/kesselheim1/mlperf_juwelsbooster}}.

\section{Large Scale AI Research at JSC}
\label{section:largescale}
\subsection{Large-Scale Deep Learning for Efficient Cross-Domain Transfer}
\label{sec:transfer}


Transfer learning was successfully employed already at the very rise of deep neural networks. Early architectures like AlexNet, OverFeat, or VGG-16 were pre-trained on ImageNet-1k, a large natural image dataset which serves as a gold standard in the visual image understanding community~\cite{Deng2009a,Russakovsky2015}. Importantly, after fine-tuning on various other target datasets, pre-trained models showed better performance compared to models just trained on target datasets from scratch~\cite{Razavian2014}. The combination of pre-training on a large generic dataset and transferring in an efficient way the pre-trained model to a specific target dataset, often much smaller in size than the one used during pre-training, has since proven itself as a viable strategy to create powerful models also for those scenarios where data is scarce~\cite{Kolesnikov2020}.


Recently, strong evidence was obtained that increasing the model size while at the same time increasing the amount of data and compute for pre-training results in very large models that have even stronger generalization and transfer capabilities~\cite{Belkin2019,Kaplan2020,Brown2020,Kolesnikov2020,Hernandez2021}. Pre-training of large transferable models requires larger datasets, like for instance ImageNet-21k~\cite{Deng2009a}, and is computationally very expensive. Therefore, to obtain such large models and benefit from their improved transferability and generalization, machines like JUWELS Booster are required to perform distributed training efficiently across multiple nodes using a large amount of GPUs.

\begin{figure}[!tb]
\begin{minipage}{0.6\linewidth}
\centering
\includegraphics[height=3cm,width=\linewidth]{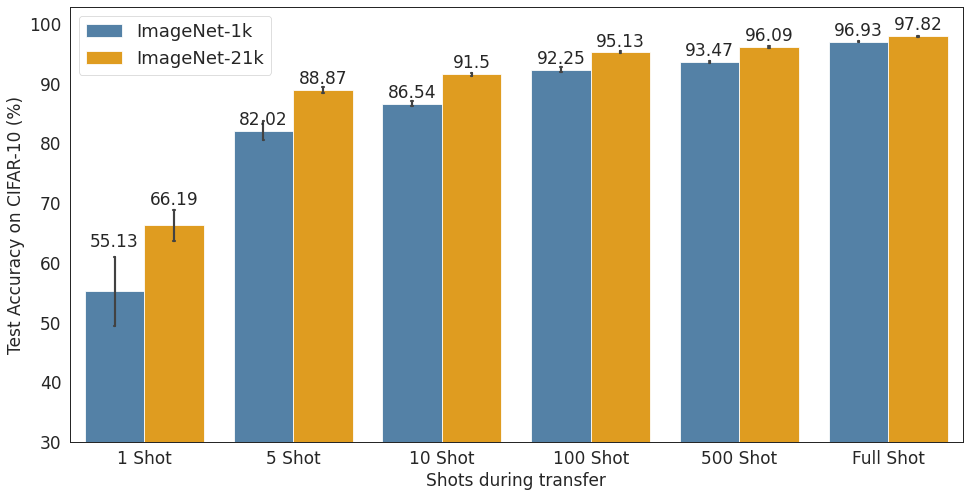}
\caption{Fine-tuning results of ResNet-152x4 on CIFAR-10 using pipeline of~\cite{Kolesnikov2020}. The model is pre-trained either on ImageNet-1k or ImageNet-21k, and we show few-shot (e.g.\ 1-shot means we use 1~example per class, thus a total of 10~examples for CIFAR-10) results as well as full fine-tuning results (the whole training set is used for fine-tuning).}
\label{fig:cifar10_finetuning}
\end{minipage}
\hspace{0.1cm}
\begin{minipage}{0.4\textwidth}
\vspace{0.5cm}
\centering
{\tiny
\begin{tabular}{@{}c|ccc|}
\hline
\multicolumn{1}{|c|}{}        & Precision & Recall & F1-score \\ \hline
\multicolumn{1}{|c|}{COVID-19}  & 0.88     &  0.84  &   0.86   \\[2.5ex] 
\multicolumn{1}{|c|}{Normal}   &  0.96    &  0.92  & 0.94      \\[2.5ex] 
\multicolumn{1}{|c|}{Pneumonia} &  0.87    & 0.93   &  0.90     \\[2.5ex] \hline
\end{tabular}}
\vspace{0.55cm}
\captionof{table}{Fine-tuning results of ResNet-152x4 on the COVIDx dataset (we use the COVIDx V7A version) from~\cite{Wang2020}. In this setup, we pre-train on ImageNet-1k and follow the fine-tuning pipeline of~\cite{Kolesnikov2020}.}
\label{table:covidx_finetuning}
\end{minipage}
\end{figure}

Here, we demonstrate the merits of such a large-scale distributed pre-training for transfer on different target datasets following~\cite{Kolesnikov2020}. We show that pre-training on ImageNet-21k  ($\approx 10$ times larger than the standard ImageNet-1k) provides a clear performance benefit in terms of accuracy achieved when transferring and fine-tuning the pre-trained model to smaller natural image datasets like CIFAR-10. Especially in the very low data regime using few-shot transfer, where only few examples per class are shown, the benefit of pre-training on large ImageNet-21k is striking (see Figure~\ref{fig:cifar10_finetuning}). JUWELS Booster shows good scaling behavior during distributed training across a vast number of compute nodes using Horovod~(Fig.~\ref{fig:benchmark}), which allows us to execute large-scale pre-training procedures in a fraction of the time for single-node training and shorten experimental cycles. Still, a full pre-training of a large ResNet-152x4 network on ImageNet-21k for 90~epochs takes ca.~81 hours when using 256~GPUs. Distributed training also performs without loss of accuracy~\cite{Shallue2019} when compared to single-node training.


Following the goals of our initiative for large-scale transfer learning applied to medical imaging for COVID-19 diagnostics (COVIDNetX\footnote{\url{https://tinyurl.com/CovidNetXHelmholtz}}), we also envisage application of large-scale generic model pre-training, for efficient and robust transfer to specific domains, like medical images. Motivated by the urgency to provide robust and widely available tools for predictive diagnostics of a patient's current and future state from medical imaging data with regard to a SARS-CoV-2 infection and its disease course~\cite{Wehbe2021,Sriram2021}, we have chosen a small publicly available dataset containing X-ray lung images of COVID-19, non-COVID, and healthy control patients (COVIDx) as a use case example for transfer~\cite{Wang2020,Cohen2020}. First preliminary results for transfer performance are available for the large ResNet-152x4 model pre-trained on ImageNet-1k (see Table~\ref{table:covidx_finetuning}). Further investigations are necessary to quantify the benefits of pre-training on much larger datasets, like ImageNet-21k, when attempting to transfer to such small, domain-specific datasets like COVIDx (see e.g.~\cite{cherti2021effect} for a follow-up study).


In summary, we have prepared the grounds for advanced transfer learning techniques, which should allow us to efficiently transfer models pre-trained on large amounts of generic data, in a highly-performant, distributed manner, to various specific target datasets of much smaller size, with low computational cost for transfer~\cite{Kolesnikov2020}. Transfer efficiency also paves the road for energy efficient deep learning that is both data-sample and compute efficient, requiring only a small energy budget for each transfer. As follow-up work, we aim to demonstrate that the postulated benefits of better generalization and transfer attributed to large-scale models~\cite{Belkin2019,Kaplan2020,Brown2020,Kolesnikov2020,Hernandez2021} hold across a very broad range of specialized domains and conditions used as transfer targets.



\subsection{Deep Learning-Driven Weather Forecast}

Weather can have an enormous effect on human lives. Improving weather forecasting can minimize the adverse effects of extreme weather and assist in planning economic activities. Numerical weather prediction~(NWP) models are among the earliest and most demanding applications for supercomputers~\cite{Bauer2015}. Recently, modern deep learning~(DL) approaches are considered to potentially play an important role in every step of the NWP workflow, from data assimilation, to replacing numerical model parameterizations, to statistical output post-processing~\cite{Bauer2015,Reichstein2019}. Because of the huge size of meteorological datasets from observations and numerical simulations, high-throughput supercomputing systems and parallel DL approaches are necessary. 

\begin{figure}[h!]
    \centering
    \includegraphics[width=0.95\linewidth]{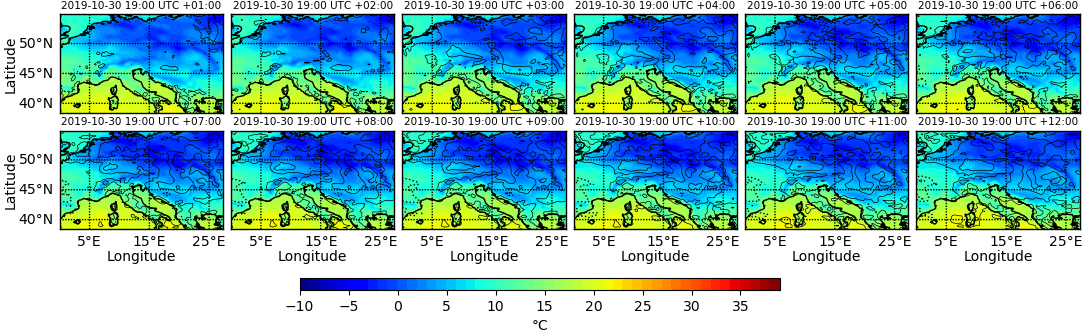} \hfill
    \caption{Example of a 2-metre temperature (\si{\celsius}) prediction  with  convLSTM (see text for details). Solid (dotted) contours denote positive (negative) temperature differences between forecast and ground truth with an interval of \SI{1}{\kelvin} (starting from \SI{0.5}{\kelvin}).
}
    \label{fig:2m_temperature_forecast}
\end{figure}

In this study, we explore the use of state-of-the-art video prediction methods to forecast meteorological variables utilizing the global ERA5 reanalysis dataset~\cite{Hersbach2020}. The goal is to forecast the 2-metre temperature over Europe for 12~hours in a data-driven way. The geographic domain consists of $56\times92$ grid points in meridional and zonal direction. Input variables are the 2-metre temperature, cloud cover, and \SI{850}{\hecto\pascal} temperatures of the preceding 12~hours. The first DL model selected is the convLSTM architecture~\cite{shi2015convolutional}. Input and output tensors of our convLSTM model have a dimension of $12 \times 56 \times 92 \times 3$ each. The model has \num{429251} parameters and was trained on 11~years of ERA5 reanalysis data in hourly resolution. The total volume of pre-processed data in TFRecords format amounts to \SI{153}{\giga\byte}. An example of a 12-hour forecast is shown in Figure~\ref{fig:2m_temperature_forecast}.

\begin{figure}[h!]
  \centering
  \begin{minipage}[h!]{0.45\textwidth}
     \includegraphics[width=.80\textwidth]{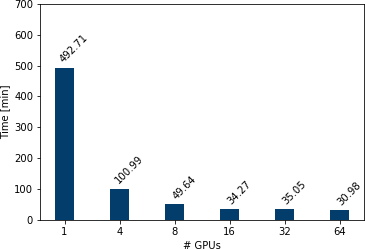}
  \end{minipage}
  \begin{minipage}[h!]{0.45\textwidth}
     \includegraphics[width=0.80\textwidth]{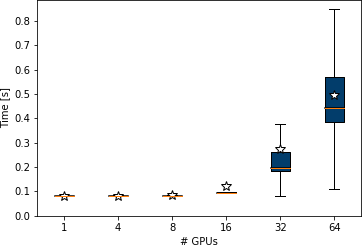}
  \end{minipage}
  \caption{Total training time in minutes~(left) and box whisker plot of iteration time in seconds~(right). The star in the box whisker plot denotes the averaged iteration time, while its median is highlighted by an orange line.}\label{fig:2m_temperature_parallel_training}
\end{figure}

Training on a single A100 GPU takes about 50~min/epoch. Given that a typical DL experiment typically requires up to 100~epochs to converge, such training times are prohibitive for any serious application, so training is parallelized using Horovod. Figure~\ref{fig:2m_temperature_parallel_training} shows that the model training achieves \SI{90}{\percent} scaling efficiency in terms of time comparing 1~GPU against 16~GPUs for 10~epochs. However, it is observed that time variances for all iterations increase significantly beyond 32~GPUs. This could be caused by data loading inefficiency, communication, or lack of enough GPU utilization. These issues are currently being investigated. After solving these issues, we plan to test more complex video prediction models (e.g.\ the stochastic adversarial video prediction architecture~\cite{lee2018stochastic}) to improve prediction accuracy and push JUWELS Booster to its limits.

\subsection{Multispectral Remote Sensing Image Classification}

Among other application areas, spaceborne Remote Sensing (RS) can detect and observe the characteristics of the Earth's surface by measuring its reflected and emitted radiation at a distance. Applications from RS use multispectral RS images to classify different physical features that occupy the surface of the Earth (i.e.\ land-cover classes), 
as well as to describe the use of the land surface by humans 
(i.e.\ land-use classes)~\cite{Canty2014}.
%
%
The RS community has started to produce several labeled RS datasets with large spatial coverage and variety of classes (e.g.\ BigEarthNet~\cite{Sumbul2019}, SEN12MS~\cite{Schmitt2019}, etc.). 
These datasets provide a high number of reliably labeled samples that can be used to train deep neural networks with supervised learning.

\begin{figure}[H]
 \centering
\subfloat[]{\includegraphics[width=0.23\linewidth]{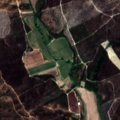}} \hfill
\subfloat[]{\includegraphics[width=0.23\linewidth]{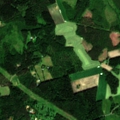}} \hfill
\subfloat[]{\includegraphics[width=0.23\linewidth]{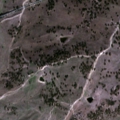}} \hfill
\subfloat[]{\includegraphics[width=0.23\linewidth]{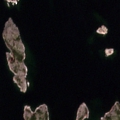}}
\caption[]{Example of patches extracted from Sentinel-2 tiles~\cite{sumbul2020bigearthnet}. The corresponding classes are: (\textbf{a})~"Permanent crops", "Broad-leaved forest", "Transitional woodland/shrub", (\textbf{b})~"Land principally occupied by agriculture, with significant areas of natural vegetation", "Broad-leaved forest", "Mixed forest", (\textbf{c})~"Pastures", "Agro-forestry areas", (\textbf{d})~"Mixed forest", "Marine waters".}
\label{fig:patches}
\end{figure}

BigEarthNet-S2\footnote{http://bigearth.net/} is a large RS archive consisting of \num{590326} patches (an example is shown in Fig.~\ref{fig:patches}) extracted from tiles acquired by the Sentinel-2 satellites~\cite{sumbul2020bigearthnet}. We train the models on the version of the dataset with 19~labels. Each patch can be associated with multiple labels, making this a multi-label classification problem. For the experiments we use the following spectral bands: 3~RGB bands and band~8 at \SI{10}{\meter} resolution, bands~5, 6, 7, 8a, 11, and~12 at \SI{20}{\meter} resolution, and bands~1 and~9 at \SI{60}{\meter} resolution. The bands at lower resolution are upsampled to \SI{10}{\meter} resolution using bilinear interpolation.


A multispectral ResNet-152 is trained from scratch on the training subset (\SI{60}{\percent} of the entire dataset) and evaluated on the test subset (\SI{20}{\percent} of the dataset).
We run the experiments with the NovoGrad optimizer. The values of the learning rate and weight decay follow the choices of~\cite{ginsburg2020stochastic}. Data augmentations with random flips, rotation of the patches, and mix-up are applied to reduce overfitting. The experiments are carried out for 100~epochs, on~1, 4, 16,~and 64~nodes (4, 16, 64~and, 256~GPUs respectively). Horovod is employed to distribute the training on multiple compute nodes and GPUs~\cite{Sergeev2018}. The batch size per GPU selected for the experiments is~16, with a global batch size that ranges from~64 for 1-node to~4096 for the 64-node configurations respectively.

The results of the classification are evaluated using the macro F1~score, which remains stable among the experiments~(0.73), and is in line with~\cite{sumbul2020bigearthnet}. Using a data-parallel approach allows us to scale up a DL model on a large remote sensing dataset and consequently cut the training time down significantly, with an \SI{80}{\percent} efficiency comparing 1~node (ca.\ \SI{2550}{\second} per epoch) against 64~nodes (ca.\ \SI{50}{\second} per epoch). Various research questions remain open, and one salient point is understanding the feasibility of training with larger global batch sizes, which are known to pose optimization difficulties~\cite{goyal2018accurate}.
A comparison between different training strategies, such as the choice of the learning rate and optimizer, is also in the future plans of the authors. More effort is also needed to enhance the pre-processing and data loading pipeline to feed the model and possibly increase efficiency when using a large number of nodes.

\subsection{RNA Structure with ML}
On a fundamental level, all life as we know it is orchestrated by interactions between biomolecules, such as proteins, DNA, and RNA. Due to a direct structure-function relationship, it is important to structurally resolve biomolecules, even though this is experimentally challenging. A complementary approach is using statistical tools to mine the rich existing protein databases. Physics-based co-evolutionary models such as direct coupling analysis~(DCA)~\cite{weigt2009identification,schug2009high,ZerihunEtalpydca2019} have in the last decade lead to the prediction of protein structures and complexes with astonishing accuracy~\cite{dago2012structural,uguzzoni2017large} by a combination of bioinformatics tools with molecular simulations. Considering the aforementioned large, public databases with \num{100000}+ biomolecular structures and even larger sequence databases, the combination of these methods with machine learning~(ML) approaches appears natural and has further improved their accuracy~\cite{Senior2020}.

Another class of biomolecules, Ribonucleic acids~(RNA), is critical for biological activities such as coding, regulation and expressions of genes. This critical importance also leads to RNA's application in pharmacology, as some of the most effective drugs in the current COVID epidemic are RNA-based. As for proteins, RNA function is closely related to its three-dimensional structure. It is, however, more challenging to gain structural information on RNA,  as reflected by the small number of RNA 3-D~structures in databases with many crucial RNA being still structurally unresolved -- akin to dark matter of the biomolecular universe~\cite{PucciAndSchug2019}. Unfortunately, while DCA still works well on RNA~\cite{DeLeonardisEtal2015,CuturelloEtal2020,PucciAndSchug2019}, ML methods so successful for proteins cannot be easily applied for RNA structure prediction, given that existing databases are considerably smaller~\cite{RFAM}. Still, even the small amount of existing data can be used to significantly improve prediction of RNA by shallow neural networks by over \SI{70}{\percent} using simple convolutional neural networks~\cite{zerihun2020coconet}. Future work will therefore focus on using the massive computing resources of JUWELS to enhance these simple CNNs by, e.g.\  using knowledge gained from proteins by transfer learning approaches. Similarly, we plan to run complex molecular dynamics simulations~(MD) on RNA, to gain more structural insight. As these simulations needs to simulate all atoms of the RNA and the surrounding solvent, they can quickly require $\mathcal{O}(10)+$ million core hours on highly parallel systems.

\section{Summary and Outlook}
In this paper we have demonstrated the hardware configuration of JUWELS Booster, currently not only the fastest supercomputer in Europe but also the most energy-efficient large-scale machine in the world. We have elucidated the key parallelization techniques that unlock this computational power for machine learning, and have shown that JUWELS Booster reaches its potential also in practical machine learning settings. 

We have indicated how the availability of such a machine can foster developments in various research fields. We have shown that large-scale pre-training can render CV applications much more data-efficient, including medical imaging for COVID-19 detection. We demonstrated that DL video prediction methods are an option for computing weather forecasts. We have seen that large-scale data parallel training enables us to train models on large corpora of multispectral satellite images and the ML-based structure predictions can help us understanding the role of RNA for biology.

All applications not only rely on raw computational power, but also on a system design and infrastructure that allows for fast parallel implementations, and a software ecosystem that gives users the possibility for rapid development, employing groundbreaking methods. The Jülich Supercomputing Centre, hosting JUWELS Booster, brings all these ingredients together, creating a state-of-the art AI landscape. 

\section*{Acknowledgements}
This work was funded by Helmholtz Association’s Initiative and Networking Fund under project number ZT-I-0003 and HelmholtzAI computing resources (HAICORE) Funding has been obtained through grants ERC-2017-ADG 787576 (IntelliAQ) and BMBF 01 IS 18O47A (DeepRain). This work was performed in the CoE RAISE and DEEP-EST projects receiving funding from EU’s Horizon 2020 Research and Innovation Framework Programme under the grant agreement no. 951733 and no. 754304 respectively. We thank ECMWF for providing ERA-5 data. The authors gratefully acknowledge the Gauss Centre for Supercomputing e.V. (www.gauss-centre.eu) for funding this work by providing computing time through the John von Neumann Institute for Computing (NIC) on the GCS Supercomputers JUWELS, JUWELS Booster at Jülich Supercomputing Centre (JSC) and  we acknowledge computing resources from the Helmholtz Data Federation. Further computing time was provided on supercomputer JUSUF in frame of offer for epidemiology research on COVID-19 by JSC.
\bibliographystyle{splncs04}
\bibliography{main}

\end{document}